\documentclass[journal]{IEEEtran}
\usepackage{amsmath, amssymb, cite, tikz, graphicx, xspace, mathrsfs, mathtools, psfrag, chemarrow, subfigure,color,soul,graphicx,mathtools}
\usepackage{kbordermatrix}
\usepackage{enumitem}
\usepackage{bbm}
\usepackage[utf8]{inputenc}
\usepackage{booktabs}
\usepackage{amscd}
\usepackage{amsmath}
\usepackage{amssymb}
\usepackage{amsthm}

\usepackage{epsfig}
\usepackage{verbatim}
\usepackage{graphicx}
\usepackage{amsthm}
\usepackage{color}
\usepackage{ multirow }
\usepackage[noend]{algpseudocode}
\usetikzlibrary{patterns}
\usetikzlibrary{angles} 
\usepackage{eucal}
\usepackage{algorithm}
\usepackage{tikz}
\usetikzlibrary{shapes, arrows, decorations.markings, arrows.meta}
\usetikzlibrary{patterns} 
\usetikzlibrary{arrows,automata,positioning,chains,calc}
\usetikzlibrary{quotes,angles}
\newtheorem{remark}{Remark}
\makeatletter
\def\BState{\State\hskip-\ALG@thistlm}
\makeatother
\newtheorem{theorem}{Theorem}
\newtheorem{lemma}[theorem]{Lemma}
\begin{document}
	\setlength{\abovecaptionskip}{-3pt}
	\setlength{\belowcaptionskip}{1pt}
	\setlength{\floatsep}{1ex}
	\setlength{\textfloatsep}{1ex}	
	\title{Version Innovation Age and Age of Incorrect Version for Monitoring Markovian Sources}
	
	\author{\author{
			\IEEEauthorblockN{Mehrdad Salimnejad\IEEEauthorrefmark{1}, 
				Marios Kountouris\IEEEauthorrefmark{2},  Anthony Ephremides \IEEEauthorrefmark{3},
				and Nikolaos Pappas\IEEEauthorrefmark{1}}
			\IEEEauthorblockA{\IEEEauthorrefmark{1}Department of Computer and Information Science, Link\"{o}ping University, Link\"{o}ping, Sweden}
			\IEEEauthorblockA{\IEEEauthorrefmark{2}Communication Systems Department, EURECOM, Sophia-Antipolis, France}
			\IEEEauthorblockA{\IEEEauthorrefmark{3}
				Electrical and Computer Engineering, University of Maryland, College Park, MD, USA}
			\{mehrdad.salimnejad, nikolaos.pappas\}@liu.se, marios.kountouris@eurecom.fr, etony@umd.edu
	}}
	
	\maketitle
	\begin{abstract}
		 In this paper, we propose two new performance metrics, coined \emph{the Version Innovation Age (VIA)} and \emph{the Age of Incorrect Version (AoIV)} for real-time monitoring of a two-state Markov process over an unreliable channel. We analyze their performance under the \emph{change-aware}, \emph{semantics-aware}, and \emph{randomized stationary} sampling and transmission policies. 
		We derive closed-form expressions for the distribution and the average of VIA, AoIV, and AoII for these policies. We then formulate and solve an optimization problem to minimize the average VIA, subject to constraints on the time-averaged sampling cost and time-averaged reconstruction error. Finally, we compare the performance of various sampling and transmission policies and identify the conditions under which each policy outperforms the others in optimizing the proposed metrics.
		
	\end{abstract}
	
	\section{Introduction}
	Timely delivery of relevant status update packets from an information source has become increasingly crucial in various real-time communication systems, in which digital components remotely monitor and control physical entities \cite{abd2019role, shreedhar2019age}. These systems require reliable and timely exchange of useful information, coupled with efficient distributed processing, to facilitate optimal decision-making in applications such as industrial automation, collaborative robotics, and autonomous transportation systems. These challenging requirements gave rise to \emph{goal-oriented semantics-empowered communications} \cite{kountouris2021semantics}, a novel paradigm that considers the usefulness, the timeliness \cite{kaul2012real, pappas2022agebook} and the innate and contextual importance of information to generate, transmit, and utilize data in time-sensitive and data-intensive communication systems. 
	Recently, a new metric called version Age of Information (AoI) has been introduced in \cite{yates2021age}, where each update at the source is considered as a new version, thus quantifying how many versions out-of-date the information on the monitor is compared to the version at the source. Several studies have considered the version AoI as a key performance metric of the timeliness of information in networks \cite{yates2021age,yates2021timely,buyukatesversion,kaswan2022timely,kaswan2022susceptibility,kaswan2022age,mitra2023age,abd2023distribution,mitra2023learning,delfani2023version,karevvanavar2023version}. The scaling of the average version AoI in gossip networks of different sizes and topologies is investigated in \cite{yates2021age,yates2021timely,buyukatesversion,kaswan2022timely,kaswan2022susceptibility,kaswan2022age,mitra2023age,abd2023distribution}. A learning-based approach to minimize the overall average version age of the worst-performing node in sparse gossip networks is employed in \cite{mitra2023learning}.  
	The authors in \cite{delfani2023version} studied the problem of minimizing the average version AoI using a Markov Decision Process (MDP) in a scenario where an energy harvesting (EH) sensor updates the gossip network via an aggregator equipped with caching capabilities. 
	The work \cite{karevvanavar2023version} considered the problem of minimizing version AoI over a fading broadcast channel, employing a non-orthogonal multiple access (NOMA) scheme.
	
	Prior work on version AoI relies on the occurrence of \emph{content change} at the source, while the destination nodes only demand the latest version of the information from the source, regardless of the \emph{content} of the information. In other words, version AoI exclusively focuses on changes occurring in the source's content, disregarding the significance and the utility of the information. Several semantics-aware metrics \cite{stamatakis2019control,stamatakis2022semantics,maatouk2020age,liu20222,pappas2021goal,tolga21SP,popovski2020semantic,PetarProc2022,jayanth23, GunduzJSAC23,cocco2023remote,salimnejad2023ICC,fountoulakis2023goal,MSalimnejadJCN2023,MSalimnejadTCOM2024} have made a step in that direction, without though investigating the evolution of versions. In this work, we propose two new semantics-aware metrics, coined \emph{Version Innovation Age}, and \emph{ Age of Incorrect Version}, which take into account both the content and the version evolution of the information source.
	
	In this paper, we examine a time-slotted communication system where a sampler performs sampling of a two-state Markov process, acting as the information source. Consequently, the transmitter sends the sample in packet form to a remote receiver over an unreliable wireless channel. A specific action is executed at the receiver based on the estimated state of the information source. 
	We propose two novel semantics-aware metrics, namely Version Innovation Age and Age of Incorrect Version, derive general expressions for the distribution of AoII and its average, and assess the system performance under change-aware, semantics-aware, and randomized stationary sampling and transmission policies. In addition, we formulate and solve a constrained optimization problem with the average Version Innovation Age as the objective function and time-averaged sampling cost and time-averaged reconstruction error as constraints.
	
	\section{System Model}
	
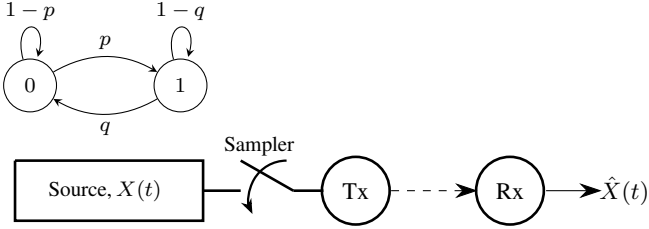
\begin{figure}[!t]
		\centering
		\footnotesize
		\begin{tikzpicture}[start chain=going left,>=stealth,node distance=2cm,on grid,auto]
			\footnotesize
			\node[state, on chain]               at (0,-2)  (2) {$1$};
			\node[state, on chain]              at  (0,-2)   (1) {$0$};
			\path[->]
			(1)   edge[loop above] node  {$1-p$}   (1)
			(1) edge  [bend left=30] node {$p$} (2)
			
			(2) edge  [bend left=30] node {$q$} (1)
			(2) edge  [loop above] node {$1-q$} (2)
			;
			\draw [line width=0.35mm](-2.2,-3)--(0.3,-3)--(0.3,-3.8)--(-2.2,-3.8)--(-2.2,-3);
			\node(a) at (-1,-3.4)  {{{Source}},\! $X(t)$};
			\draw [line width=0.35mm](0.3,-3.4) -- (0.8,-3.4);
			\draw [line width=0.35mm](0.8,-3)to(1.5,-3.4);
			\draw [line width=0.35mm](1.5,-3.4)to(1.9,-3.4);
			
			\draw[line width=0.3mm,-{Stealth[length=2mm]}] (1.41,-3) arc (90:205:5mm) ;
			\draw [line width=0.35mm](2.33,-3.4) circle (0.45cm);
			\node(b) at (2.33,-3.4)  {\small	$\text{Tx}$};
			\draw [-{Stealth[length=3mm, width=2mm]},dashed](2.8,-3.4) -- (3.95,-3.4);
			\draw [line width=0.35mm](4.38,-3.4) circle (0.45cm);
			\node(c) at (4.38,-3.4)  {\small	$\text{Rx}$};
			\draw [-{Stealth[length=3mm, width=2mm]}](4.86,-3.4) -- (5.6,-3.4);
			\node(d) at (5.9,-3.4)  {\small$\hat{X}(t)$};
			\node(e) at (1.,-2.8)  {\footnotesize$\text{ Sampler}$};
		\end{tikzpicture}
		\caption{Real-time monitoring of a Markovian source over a wireless channel.}
		\label{system_model_fig}
	\end{figure}

	We consider a time-slotted communication system in which a sampler conducts sampling of an information source, denoted as $X(t)$, at time slot $t$, as shown in Fig. \ref{system_model_fig}. The transmitter then forwards the sampled information in packets over a wireless communication channel to the receiver. The information source is represented as a two-state discrete-time Markov chain (DTMC) $\{X(t), t \in \mathbb{N}\}$. Therein, the state transition probability $\mathrm{Pr}\big[X(t+1)=j \big|X(t)=i\big]$ represents the probability of transitioning from state $i$ to $j$ and can be defined as $\mathrm{Pr}\big[X(t+1)=j\big|X(t)=i\big] = \mathbbm{1} (i=0,j=0)(1-p)+\mathbbm{1}(i=0,j=1)p+\mathbbm{1}(i=1,j=0)q+\mathbbm{1}(i=1,j=1)(1-q)$, where $\mathbbm {1}(\cdot)$ is the indicator function. We denote the action of sampling at time slot $t$ by $\alpha^{\text{s}}(t)=\{0,1\}$, where $\alpha^{\text{s}}(t)=1$ if the source is sampled and $\alpha^{\text{s}}(t)=0$ otherwise\footnote{We assume that when sampling occurs at time slot $t$, the transmitter sends the sample immediately at that time slot.}. Here, for the sampling and transmission policy, we consider \emph{the randomized stationary policy} where a new sample is performed probabilistically at each time slot \cite{MSalimnejadTCOM2024}. We assume that the probability of sampling at time slot $t$ is $p_{\alpha^{\text{s}}}=\mathrm{Pr}[\alpha^{\text{s}}(t)=1]$. We also define $\mathrm{Pr}[\alpha^{\text{s}}(t)=0]=1-p_{\alpha^{\text{s}}}$ as the probability that the source is not sampled at time slot $t$. For comparison purposes, we adopt two other relevant sampling policies, namely \emph{change-aware} and \emph{semantics-aware} policies proposed in \cite{pappas2021goal, MSalimnejadTCOM2024}. At time slot $t$, the receiver constructs an estimate of the process $X(t)$, denoted by $\hat{X}(t)$, based on the successfully received samples. 
	
	\label{system_model}
	
	We consider a wireless channel between the source and the receiver, assuming that the channel state $h(t)$ equals $1$ if the information source is sampled and successfully decoded by the receiver and $0$ otherwise. We define the success probability as $p_{s}=\mathrm{Pr}[h(t)=1]$. At time slot $t$, when the source is sampled and transmitted, we assume that with probability $p_{s}$, the system is in the sync state, i.e., $X(t)=\hat{X}(t)$. Otherwise, if the system is in an erroneous state, the state of the reconstructed source remains unchanged, i.e., $\hat{X}(t)=\hat{X}(t-1)$. Acknowledgment (ACK)/negative-ACK(NACK) packets are used to inform the transmitter about the success or failure of transmissions. ACK/NACK packets are assumed to be delivered to the transmitter instantly and without errors\footnote{An ACK/NACK feedback channel is required only for the semantics-aware policy.}. Therefore, the transmitter has accurate information about the reconstructed source state at time slot $t$, i.e., $\hat{X}(t)$. In addition, we assume that the corresponding sample is discarded in the event of a transmission failure (packet drop channel). 
		\section{Performance Metrics}
	In this section, we consider the impact of the semantics of information at the receiver. We propose and analyze two new performance metrics, termed \emph{Version Innovation Age} (VIA) and \emph{Age of Incorrect Version} (AoIV), which jointly quantify \emph{both the timing and importance aspects of information}. Furthermore, we compare the average Age of Incorrect Information (AoII) under the policies presented in Section \ref{system_model}.
	
	\subsection{Version Innovation Age (VIA)}
	\par Before introducing the new metric in this section, we first review the Version Age of Information (VAoI) proposed in \cite{yates2021age}. We assume that each update at the information source represents a version. At time slot $t$, $V_{\text{S}}(t)$ represents the version at the source, whilst $V_{\text{R}}(t)$ represents the version at the receiver. The VAoI at the receiver is then defined as $\text{VAoI}(t) = V_{\text{R}}(t)-V_{\text{S}}(t)$. This metric relies on changes in the information source, but the content of the information source is not considered important. Therefore,  we introduce here a new metric named the Version Innovation Age (VIA), which measures the number of outdated versions at the receiver compared to the source when the source is in a specific state. We define the evolution of VIA as follows
	\begin{align}
		\label{Version_AoI}
		{\text{VIA}}\big(t\!+\!1\big)\!\!=\!\!
		\begin{cases}
			\!\!{\text{VIA}}(t), &\!\!\!\!\parbox[t]{6cm}{{$X(t)\!=\!X(t\!+\!1)$ \!\text{and} $\{\alpha^{\text{s}}(t)\!=\!0,\\ \text{or}\hspace{0.1cm}(\alpha^{\text{s}}(t)\!=\!1, h(t)\!=\!0)\}$,}\\}\\
			\!\!{\text{VIA}}(t)\!+\!1, & \!\!\!\!\parbox[t]{9cm}{{$X(t)\!\neq\! X(t\!+\!1)$ \!\text{and} $\{\alpha^{\text{s}}(t)\!=\!0,\\ \text{or} \hspace{0.1cm} (\alpha^{\text{s}}(t)\!=\!1, h(t)\!=\!0)\}$,}\\}\\
			0, &\!\!\! \parbox[t]{9cm}{$h(t)=1$.}\\
		\end{cases}
	\end{align}
	\begin{figure}[!t]
		\centering
		\begin{tikzpicture}[start chain=going left,->,>=latex,node distance=2cm,on grid,auto]
			\footnotesize
			\node[on chain]                        (g) {$\cdots$};
			\node[state, on chain]                 (3) {$2$};
			\node[state, on chain]                 (2) {$1$};
			\node[state, on chain]                 (1) {$0$};
			\draw[>=latex]
			(1)   edge[loop above] node {$P_{0,0}$}   (1)
			(1) edge  [bend left=30] node {$P_{0,1}$} (2)
			
			(2) edge  [bend left=30] node {$P_{1,2}$} (3)
			(2) edge  [bend left=30] node[above] {$P_{1,0}$} (1)
			(2) edge  [loop above] node {$P_{1,1}$} (2)
			
			(3) edge  [bend left=40] node {$P_{2,0}$} (1)
			(3) edge  [bend left=30] node {$P_{2,3}$} (g)
			(3) edge  [loop above] node {$P_{2,2}$} (2)
			
			(g) edge  [bend left=50] node {$P_{3,0}$} (1)
			;
		\end{tikzpicture}
		\caption{DTMC describing the evolution of the VIA.}
		\label{AoIV_DTMC}
	\end{figure}
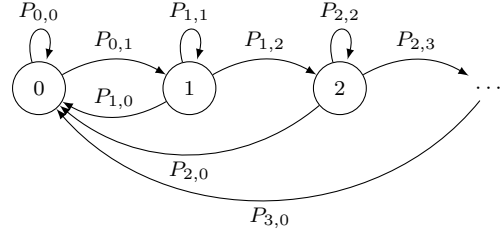
	Using \eqref{Version_AoI}, we can describe this metric by a DTMC as depicted in Fig. \ref{AoIV_DTMC}. The transition probability ${\text{VIA}}(t)$ is now defined as 
	\begin{align}
		\label{Pij_AoIV}
		P_{i,j} = \mathrm{Pr}\big[{\text{VIA}}(t+1)=j|{\text{VIA}}(t)=i\big].
	\end{align}
	\begin{lemma}
		\label{Lemma_Pij}
		For the randomized stationary policy, the transition probability $P_{i,j}, \forall i,j\in \{0,1,\cdots\}$ is given by
		\begin{align}
			\label{Pij_lemma}
			P_{i,j} &= \Big[\mathbbm{1}(j\!=\!0)p_{\alpha^{\text{s}}}p_{\text{s}}+\mathbbm{1}(j\!=\!i)(1-p)(1-p_{\alpha^{\text{s}}}p_{\text{s}})\notag\\
			&+\mathbbm{1}(j\!=\!i+1)p(1-p_{\alpha^{\text{s}}}p_{\text{s}})\Big]\frac{\pi_{0,i}}{\pi_{0,i}+\pi_{1,i}}\notag\\
			&+\Big[\mathbbm{1}(j\!=\!0)p_{\alpha^{\text{s}}}p_{\text{s}}+\mathbbm{1}(j\!=\!i)(1-q)(1-p_{\alpha^{\text{s}}}p_{\text{s}})\notag\\
			&+\mathbbm{1}(j\!=\!i+1)q(1-p_{\alpha^{\text{s}}}p_{\text{s}})\Big]\frac{\pi_{1,i}}{\pi_{0,i}+\pi_{1,i}}.
		\end{align}
		Note that $\pi_{0,i}$ and $\pi_{1,i}$ in Lemma \ref{Lemma_Pij} are the probabilities obtained from the stationary distribution of the two-dimensional DTMC describing the joint status of the original source regarding the current state of the VIA, i.e., $\big(X(t), {\text{VIA}}(t)\big)$.
	\end{lemma}
	\begin{IEEEproof} 
		See Appendix \ref{Appendix_LemmaPij}.
	\end{IEEEproof}
	
	\begin{lemma}
		\label{theorem_VAoI}
		For a two-state DTMC information source, the stationary distribution $\pi_{j,i}$ for the randomized stationary policy is given by
		\begin{align}
			\!\!\pi_{0,i}\!\!&=\!\!\frac{p^kq^wp_{\alpha^{\text{s}}}p_{s}\big(1\!-\!p_{\alpha^{\text{s}}}p_{s}\big)^i}{(p\!+\!q)\big(p\!+\!(1\!-\!p)p_{\alpha^{\text{s}}}p_{s}\big)^w\big(q\!+\!(1\!-\!q)p_{\alpha^{\text{s}}}p_{s}\big)^k},
			i\!=\!0,1,\ldots.\notag\\
			\!\!\pi_{1,i}\!\!&=\!\!\frac{p^wq^kp_{\alpha^{\text{s}}}p_{s}\big(1\!-\!p_{\alpha^{\text{s}}}p_{s}\big)^i}{(p\!+\!q)\big(p\!+\!(1\!-\!p)p_{\alpha^{\text{s}}}p_{s}\big)^k\big(q\!+\!(1\!-\!q)p_{\alpha^{\text{s}}}p_{s}\big)^w}, i\!=\!0,1,\ldots.\label{pi0i_pi1i_AoI_RS}
		\end{align}
		where $k$, and $w$ are given by
		\begin{align}
			\label{kw_AoI}
			k=
			\begin{cases}
				\frac{i}{2}, &\mod\{i,2\}=0,\\
				\frac{i+1}{2}, &\mod\{i,2\}\neq 0.
			\end{cases}\notag\\
			\hspace{0.2cm}w=
			\begin{cases}
				\frac{i+2}{2}, &\mod\{i,2\}=0,\\
				\frac{i+1}{2}, &\mod\{i,2\}\neq 0.
			\end{cases}
		\end{align}
	\end{lemma}
	
	\begin{IEEEproof} 
		See Appendix \ref{Appendix_LemmaPij_VAoI}.
	\end{IEEEproof}
	
	Using \eqref{pi0i_pi1i_AoI_RS} and \eqref{kw_AoI}, we can calculate the average VIA, $\overline{\text{VIA}}$, as
	\begin{align}
		\label{Avg_AoIV_RS}
		\overline{\text{VIA}} &= \sum_{i=0}^{\infty} i\mathrm{Pr}[{\text{VIA}}(t)=i] =  \sum_{i=1}^{\infty} i\big(\pi_{0,i}+\pi_{1,i}\big)\notag\\
		&=\frac{2pq\big(1-p_{\alpha^{\text{s}}}p_{s}\big)}{(p+q)p_{\alpha^{\text{s}}}p_{s}}.
	\end{align}
	Note that the convergence condition for the previous expression is $\!\!\frac{\sqrt{pq}\big(1\!-\!p_{\alpha^{\text{s}}}p_{s}\big)}{\sqrt{\!\big(p\!+\!(1-p)p_{\alpha^{\text{s}}}p_{s}\big)\!\!\big(q\!+\!(1-q)p_{\alpha^{\text{s}}}p_{s}\big)}}\!\!<\!\!1$.
	
	\begin{remark}
		\label{remark_RS_CA_compare_AoIV}
		Using \eqref{Avg_AoIV_RS} and \eqref{Avg_AoIV_CA}, we can prove that when $p_{\text{s}}=1$ or $2pq=p+q$, or $p_{\text{s}}=0$ and $2pq=p+q$, the randomized stationary policy has the same average VIA as the change-aware policy only if $p_{\alpha^{\text{s}}}=1$; otherwise, the change-aware policy has a lower average VIA compared to the randomized stationary policy. Furthermore, when $0<p_{\text{s}}<1$ and $2pq\neq p+q$, the randomized stationary policy has a lower average VIA compared to the change-aware policy if $\frac{2pq}{p+q+\big(2pq-p-q\big)p_{\text{s}}}\leqslant p_{\alpha^{\text{s}}}<1$.
	\end{remark}
	\begin{remark}
		\label{remark_RS_CA_AoIV_PE}
		Using eq. (39) in \cite{MSalimnejadTCOM2024}, we can express the average VIA given in \eqref{Avg_AoIV_RS} and \eqref{Avg_AoIV_CA} as a function of the time-averaged reconstruction error. For the randomized stationary policy, \eqref{Avg_AoIV_RS} is formulated as
		\begin{align}
			\label{PE_AoIV_RS}
			\overline{\text{VIA}}(P_{E})=\frac{\big[p+q+(1-p-q)p_{\alpha^{\text{s}}}p_{\text{s}}\big]P_{E}}{p_{\alpha^{\text{s}}}p_{\text{s}}}, 
		\end{align}
		where $P_{E}$ represents the time-averaged reconstruction error and in \eqref{PE_AoIV_RS} is given by
		\begin{align}
			\label{PE_RS}
			P_{E} = \frac{2pq(1-p_{\alpha^{\text{s}}}p_{\text{s}})}{(p+q)\big[p+q+(1-p-q)p_{\alpha^{\text{s}}}p_{\text{s}}\big]}.
		\end{align}
		Furthermore, for the change-aware policy, the expression given in \eqref{Avg_AoIV_CA} can be written as 
		\begin{align}
			\label{PE_AoIV_CA}
			\overline{\text{VIA}}(P_{E}) = \bigg(\frac{2}{p_{\text{s}}}-1\bigg)P_{E},
		\end{align}
		where $P_{E}$ in \eqref{PE_AoIV_CA} is calculated as
		\begin{align}
			\label{PE_CA}
			P_{E}=\frac{1-p_{\text{s}}}{2-p_{\text{s}}}.
		\end{align}
	\end{remark}
	
	\subsection{Age of Incorrect Version (AoIV)}
	\par A major issue with VIA is that when the state of the source changes and transmission fails, the VIA increases by one. However, the system may be in a synced state, i.e., $X(t) = \hat{X}(t)$. In other words, even if the system has perfect knowledge of the source's state, the VIA can still increase. For that, we introduce another metric, named Age of Incorrect Version (AoIV), which is defined as the number of outdated versions at the receiver compared to the source when the system is in an erroneous state, i.e., $X(t) \neq \hat{X}(t)$. We can define the evolution of the AoIV as follows
	\begin{align}
		\label{Version_AoII}
		{\text{AoIV}}\big(t+1\big) =
		\begin{cases}
			{\text{AoIV}}(t),&\parbox[t]{9cm}{{$X(t+1)=X(t),\\X(t+1)\neq\hat{X}(t+1)$}\\}\\ 
			{\text{AoIV}}(t)+1,&\parbox[t]{9cm}{{$X(t+1)\neq X(t),\\X(t+1)\neq\hat{X}(t+1)$}\\}\\ 
			0, &X(t+1)=\hat{X}(t+1).
		\end{cases}
	\end{align}
	Using \eqref{Version_AoII}, for a two-state DTMC information source, $\mathrm{Pr}[{\text{AoIV}}(t)\neq 0]$ is calculated as
	\begin{align}
		\label{PrDAoIVI}
		\mathrm{Pr}[{\text{AoIV}}(t)\neq 0] &= \mathrm{Pr}\big[X(t)=0,\hat{X}(t)=1,{\text{AoIV}}(t)=1\big]\notag\\&+\mathrm{Pr}\big[X(t)=1,\hat{X}(t)=0,{\text{AoIV}}(t)=1\big]\notag\\&=\pi_{0,1,1}+\pi_{1,0,1},
	\end{align}
	where $\pi_{0,1,1}$ and $\pi_{1,0,1}$ are the probabilities obtained from the stationary distribution of the three-dimensional DTMC describing the joint status of the original and reconstructed source regarding the current state of the AoIV, i.e., $\big(X(t),\hat{X}(t),{\text{AoIV}}(t)\big)$. 
	\begin{lemma}
		\label{piijk_VAoII_RS}
		For a two-state DTMC information source, the stationary distribution $\pi_{i,j,k}, \forall i,j,k \in\{0,1\}$ for the randomized stationary policy is given by\footnote{For a two-state DTMC information source, the maximum value of AoIV is equal to 1.}
		
		\begin{subequations}
			\label{VAoII_pi_ijk_RS}
			\begin{align}
				\pi_{0,0,0}
				&=\frac{q\big[q+(1-q)p_{\alpha^{\text{s}}}p_{s}\big]}{(p+q)\big[p+q+(1-p-q)p_{\alpha^{\text{s}}}p_{s}\big]},\label{pi000_1}\\
				\pi_{0,1,1}
				&=\frac{pq\big(1-p_{\alpha^{\text{s}}}p_{s}\big)}{(p+q)\big[p+q+(1-p-q)p_{\alpha^{\text{s}}}p_{s}\big]},\label{pi011_1}\\
				\pi_{1,1,0}
				&=\frac{p\big[p+(1-p)p_{\alpha^{\text{s}}}p_{s}\big]}{(p+q)\big[p+q+(1-p-q)p_{\alpha^{\text{s}}}p_{s}\big]},\label{pi110_1}\\
				\pi_{1,0,1}
				&=\frac{pq\big(1-p_{\alpha^{\text{s}}}p_{s}\big)}{(p+q)\big[p+q+(1-p-q)p_{\alpha^{\text{s}}}p_{s}\big]}\label{pi101_1},\\
				\pi_{0,0,1}&=\pi_{0,1,0}=\pi_{1,0,0}=\pi_{1,1,1}=0.
			\end{align}
		\end{subequations}
	\end{lemma}
	\begin{IEEEproof}
		See Appendix \ref{Appendix_piijk_VAoII_RS}.
	\end{IEEEproof} 
	
	Now, using \eqref{pi011_1} and \eqref{pi101_1} we can calculate \eqref{PrDAoIVI} as follows
	\begin{align}
		\label{PrDAoIVI2}
		\!\!\!	\mathrm{Pr}[{\text{AoIV}}(t)=1]&=\pi_{0,1,1}+\pi_{1,0,1}\notag\\
		&=\frac{2pq\big(1-p_{\alpha^{\text{s}}}p_{s}\big)}{(p+q)\big[p+q+(1-p-q)p_{\alpha^{\text{s}}}p_{s}\big]}.
	\end{align}
	Using \eqref{PrDAoIVI2}, the average AoIV, $\overline{\text{AoIV}}$, can be obtained as
	\begin{align}
		\label{Avg_AoIVI_RS}
		\overline{\text{AoIV}} &= \sum_{i=1}^{\infty} i\mathrm{Pr}[{\text{AoIV}}(t)=i]\notag\\& =  \frac{2pq\big(1-p_{\alpha^{\text{s}}}p_{s}\big)}{(p+q)\big[p+q+(1-p-q)p_{\alpha^{\text{s}}}p_{s}\big]}.
	\end{align}
	
	\subsection{Age of Incorrect Information}
	\par The \emph{Age of Incorrect Information (AoII)} is a metric that quantifies the time elapsed since the transmitted information became incorrect or outdated \cite{maatouk2020age}. Let $\text{AoII}(t)\neq 0$ denote the system being in an erroneous state, i.e., $X(t)\neq \hat{X}(t)$, while the synced state of the system is denoted by $\text{AoII}(t)=0$. We also define $\text{AoII}(t)$ as the AoII at time slot $t$. The evolution of this metric can be described as follows
	\begin{align}
		\label{AoII_definition}
		\text{AoII}\big(t+1\big) =
		\begin{cases}
			\text{AoII}(t)+1, &X(t+1)\neq \hat{X}(t+1),\\
			0, &\text{otherwise}.
		\end{cases}
	\end{align}
	\begin{lemma}
		\label{AoIIDistribution_RS}
		For a two-state DTMC information source, $\mathrm{Pr}[\text{AoII}(t)=i]$ for the randomized stationary policy is given by
		\begin{align}
			&\mathrm{Pr}[\text{AoII}(t)=i] \notag\\
			&= 
			\begin{cases}
				\frac{p^2+q^2+(p+q-p^2-q^2)p_{\alpha^{\text{s}}}p_{\text{s}}}{(p+q)\big[p+q+(1-p-q)p_{\alpha^{\text{s}}}p_{\text{s}}\big]},\hspace{0.2cm} &i=0,\\
				\frac{pq(1-p_{\alpha^{\text{s}}}p_{\text{s}})^{i}\big[(1-q)^{i-1}\Phi(q)+(1-p)^{i-1}\Phi(p)\big]}{(p+q)\big[p+q+(1-p-q)p_{\alpha^{\text{s}}}p_{\text{s}}\big]},\hspace{0.2cm} &i\geqslant 1.
			\end{cases}
		\end{align}
		where $\Phi(\cdot)$ is given by
		\begin{align}
			\label{Phix}
			\Phi(x) = x+(1-x)p_{\alpha^{\text{s}}}p_{\text{s}}.
		\end{align} 
	\end{lemma}
	\begin{IEEEproof}
		See Appendix \ref{Appendix_AoIIDistribution_RS}.    
	\end{IEEEproof}
	Using Lemma \ref{AoIIDistribution_RS}, we can calculate the average AoII $\overline{\text{AoII}}$ as follows
	\begin{align}
		\label{Avg_AoII_RS}
		\overline{\text{AoII}} &=\sum_{i=1}^{\infty} i\mathrm{Pr}[\text{AoII}(t)\!=\!i]\notag\\ &=\frac{pq(1-p_{\alpha^{\text{s}}}p_{s})\big[p+q+(2-p-q)p_{\alpha^{\text{s}}}p_{s}\big]}{(p+q)\Phi(p)\Phi(q)\big[p+q+(1-p-q)p_{\alpha^{\text{s}}}p_{s}\big]}
	\end{align}
	where $\Phi(\cdot)$ is obtained in \eqref{Phix}.  
	
	\section{Optimization Problem}
	\par In this section, we aim to find the optimal randomized stationary sampling policy that minimizes the average VIA, while considering constraints on both the time-averaged sampling cost and the time-averaged reconstruction error given in \eqref{PE_RS}. We assume that each attempted sampling incurs a cost denoted by $\delta$, and the time-averaged sampling cost is constrained not to surpass a specified threshold $\delta_{\text{max}}$. Furthermore, we consider a constraint that the time-averaged reconstruction error cannot exceed a certain threshold $E_{\text{max}}$. We formulate the following optimization problem 
	\begin{subequations}
		\label{Optimization_problem1}
		\begin{align}
			&\underset{p_{\alpha^{\text{s}}}}{\text{minimize}}\hspace{0.3cm}\overline{\text{VIA}}\label{Optimization_prob1_objfunc}\\
			&\text{subject to}\hspace{0.2cm} \lim_{T \to \infty}\frac{1}{T}\sum_{t=1}^{T}\delta \mathbbm{1}\{\alpha^{\text{s}}(t)=1\} \leqslant\delta_{\text{max}},\label{Optimization_prob1_constraint1}\\
			&\hspace{1.8cm}	P_{\text{E}}\leqslant E_{\text{max}},\label{Optimization_prob1_constraint2}
		\end{align}
	\end{subequations}
	where the constraint in \eqref{Optimization_prob1_constraint1} is the time-averaged sampling cost, which can be simplified as
	\begin{align}
		\label{sampling_cosnt}
		\lim_{T \to \infty}\frac{1}{T}\sum_{t=1}^{T}\delta \mathbbm{1}\{\alpha^{\text{s}}(t)=1\}=\delta p_{\alpha^{\text{s}}}.
	\end{align}
	Now, using \eqref{Avg_AoIV_RS}, \eqref{PE_RS} and \eqref{sampling_cosnt} the optimization problem can be formulated as
	\begin{subequations}
		\label{Optimization_problem}
		\begin{align}
			&\!\!\!\!\underset{p_{\alpha^{\text{s}}}}{\text{minimize}}\hspace{0.3cm}\frac{2pq\big(1-p_{\alpha^{\text{s}}}p_{s}\big)}{(p+q)p_{\alpha^{\text{s}}}p_{s}}\label{Optimization_prob_objfunc}\\
			&\!\!\!\!\text{subject to}\hspace{0.2cm}
			\frac{2pq-E_{\text{max}}(p+q)^2}{2pqp_{\text{s}}+E_{\text{max}}(p+q)(1-p-q)p_{\text{s}}}\leqslant p_{\alpha^{\text{s}}}\leqslant \eta,\label{Optimization_prob_constraint}
		\end{align}
	\end{subequations}
	where $\eta = \delta_{\text{max}}/\delta\leqslant 1$ and $ E_{\text{max}}\leqslant 1$.
	\par To solve this optimization problem, we first note that the objective function in \eqref{Optimization_prob_objfunc} is decreasing with $p_{\alpha^{\text{s}}}$. In other words, the objective function has its minimum value when $p_{\alpha^{\text{s}}}$ has its maximum. Now, using the constraint given in \eqref{Optimization_prob_constraint}, the maximum value of sampling probability is $\eta$. However, $\eta$ is the optimal value of sampling probability when $ \frac{2pq-E_{\text{max}}(p+q)^2}{2pqp_{\text{s}}+E_{\text{max}}(p+q)(1-p-q)p_{\text{s}}}\leqslant\eta\leqslant 1$; otherwise, we cannot find a sampling probability that satisfies the constraint of the optimization problem, and thus, an optimal solution does not exist.
	\begin{remark}
		In what follows, RS and RSC policies represent the randomized stationary policy and the randomized stationary policy in the constrained optimization problem, respectively.
	\end{remark}
	
	\section{Numerical Results}
	\par In this section, we numerically validate our analytical results and assess the performance of the sampling policies in terms of the average VIA and the average AoIV under various system parameters. Simulation results are obtained by averaging over $10^7$ time slots.
	\begin{figure}[!t]
		\centering
		\subfigure[$p_{{\text{s}}} = 0.3$ ]{\includegraphics[width=0.48\linewidth, clip]{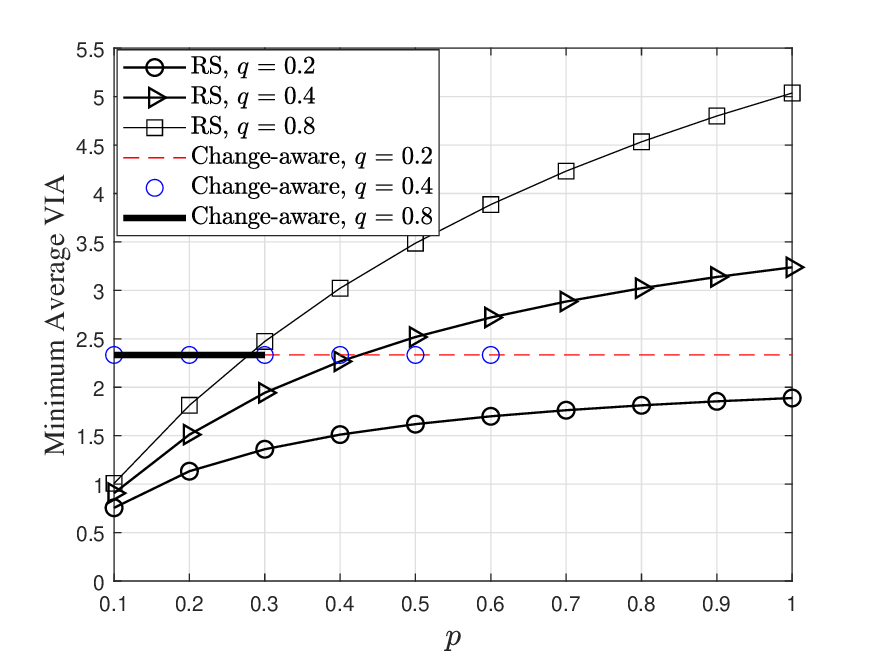}
		}
		\subfigure[$p_{{\text{s}}} = 0.7$]{\centering
			\includegraphics[width=0.48\linewidth, clip]{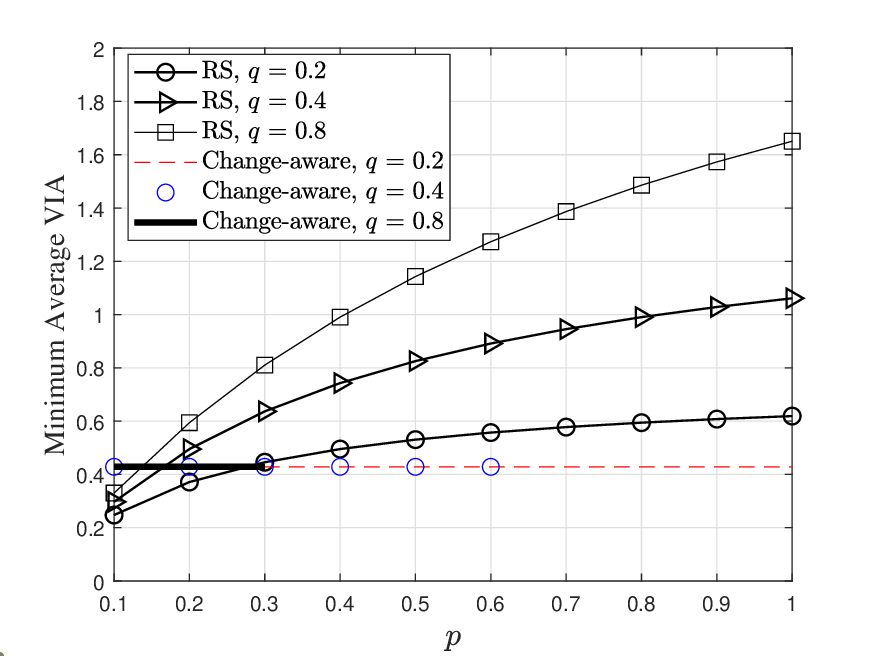}
		}
		\caption{Minimum average VIA in a constrained optimization problem as a function of $p$ and $q$.}
		\label{Min_VAoI_Constrained}
	\end{figure}
	\begin{figure}[!t]
		\centering
		\subfigure[$p_{{\text{s}}} = 0.3$ ]{\includegraphics[width=0.48\linewidth, clip]{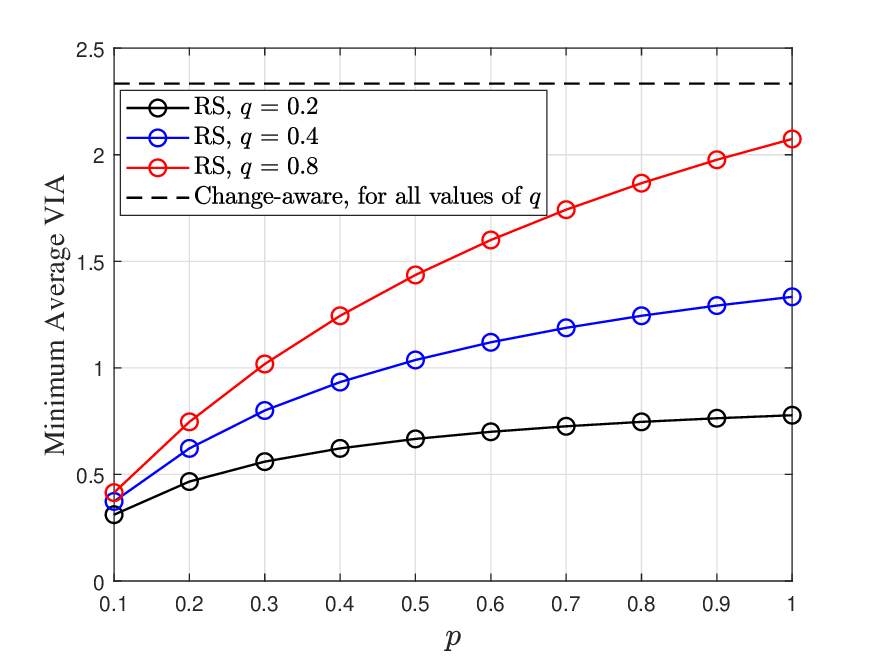}
		}
		\subfigure[$p_{{\text{s}}} = 0.7$]{\centering
			\includegraphics[width=0.48\linewidth, clip]{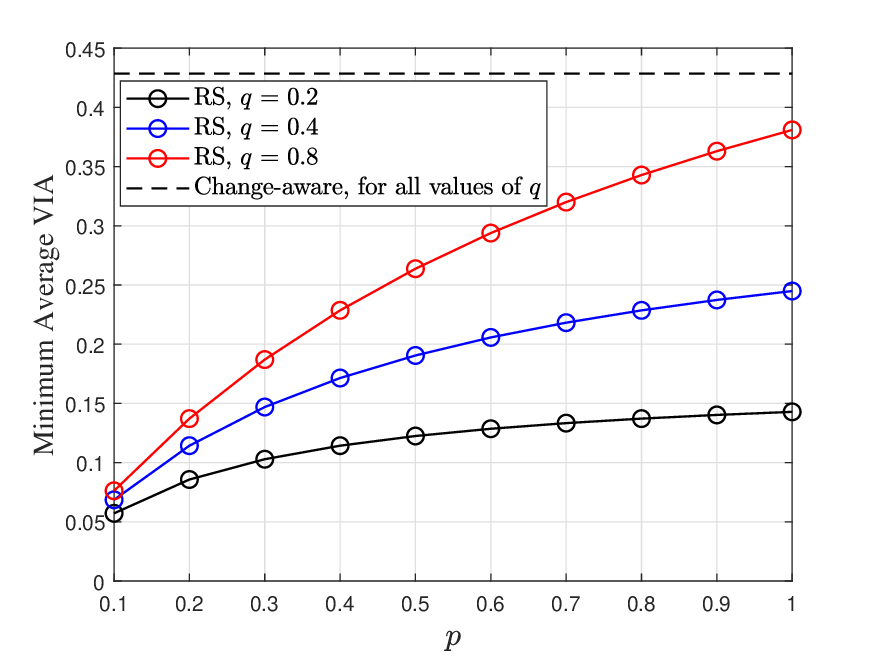}
		}
		\caption{Minimum average VIA in an unconstrained optimization problem as a function of $p$ and $q$.}
		\label{Min_VAoI_UnConstrained}
	\end{figure}
	
	Figs. \ref{Min_VAoI_Constrained} and \ref{Min_VAoI_UnConstrained} illustrate the minimum average VIA under time-averaged sampling cost and reconstruction error constraints for $\eta = 0.5$ and $E_{\text{max}} = 0.5$, considering various values of $p$, $q$, and $p_{\text{s}}$ for both the constrained and unconstrained optimization problems, respectively. As seen in Fig. \ref{Min_VAoI_Constrained}, with both low and high success probabilities, the optimal RS outperforms the change-aware policy in scenarios where the source changes slowly and rapidly. In contrast, the change-aware policy exhibits superior performance for a moderately changing source. This is because when the source changes slowly, the change-aware policy takes fewer sampling and transmission actions, resulting in a higher average VIA. In contrast, when the source changes rapidly, the change-aware policy generates more samples, violating the imposed constraint, and in that case, the optimal RS performs better. Moreover, according to Remark \ref{remark_RS_CA_compare_AoIV}, the randomized stationary policy achieves a lower average VIA compared to the change-aware policy when $\frac{2pq}{p+q+\big(2pq-p-q\big)p_{\text{s}}}\leqslant p_{\alpha^{\text{s}}}<1$. However, based on the constraint given in \eqref{Optimization_prob_constraint}, the maximum sampling probability is limited to $\eta$. Consequently, for values of $p$ and $q$ where $\eta<\frac{2pq}{p+q+\big(2pq-p-q\big)p_{\text{s}}}$, the change-aware policy outperforms the optimal RS policy in terms of average VIA. Furthermore, as shown in Fig. \ref{Min_VAoI_UnConstrained}, the optimal performance of the RS policy surpasses that of the change-aware policy in the unconstrained case. However, in such a scenario, the optimal strategy for the RS policy involves sampling. 
	Moreover, transmitting at every time slot results in the generation of an excessive amount of samples.  
	\begin{figure}[!t]
		\centering
		\subfigure[$p_{{\text{s}}} = 0.3$ ]{\includegraphics[width=0.48\linewidth, clip]{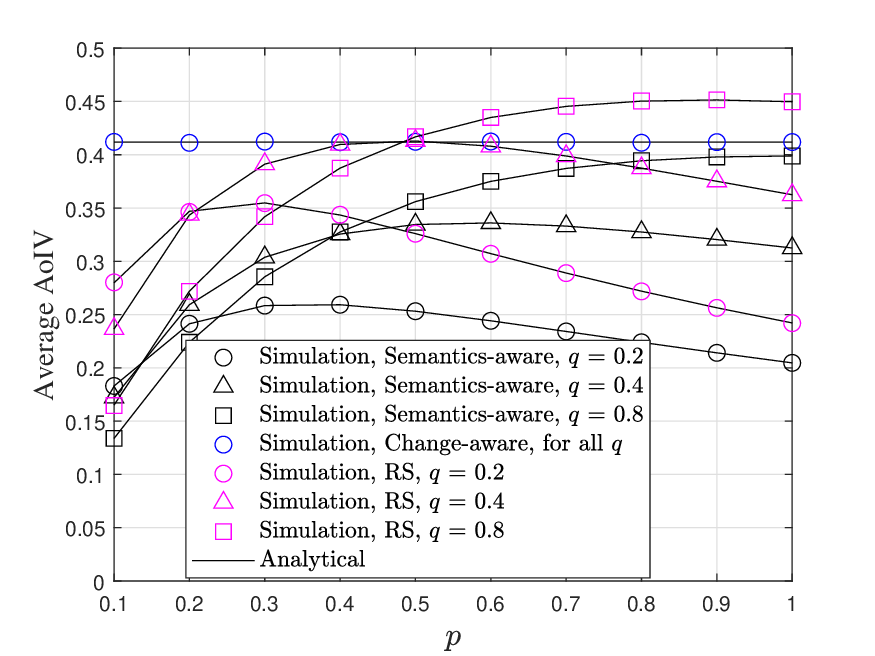}
		}
		\subfigure[$p_{{\text{s}}} = 0.7$]{\centering
			\includegraphics[width=0.48\linewidth, clip]{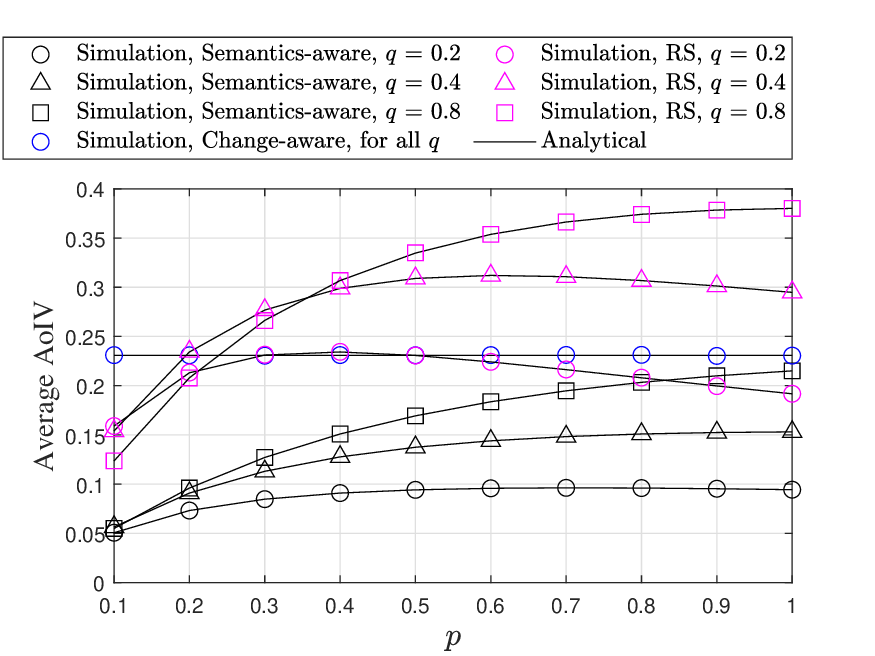}
		}
		\caption{Average AoIV as a function of $p$ and $q$.}
		\label{Avg_VAoII}
	\end{figure}
	\begin{figure}[!t]
		\centering
		\subfigure[$p_{{\text{s}}} = 0.3$ ]{\includegraphics[width=0.48\linewidth, clip]{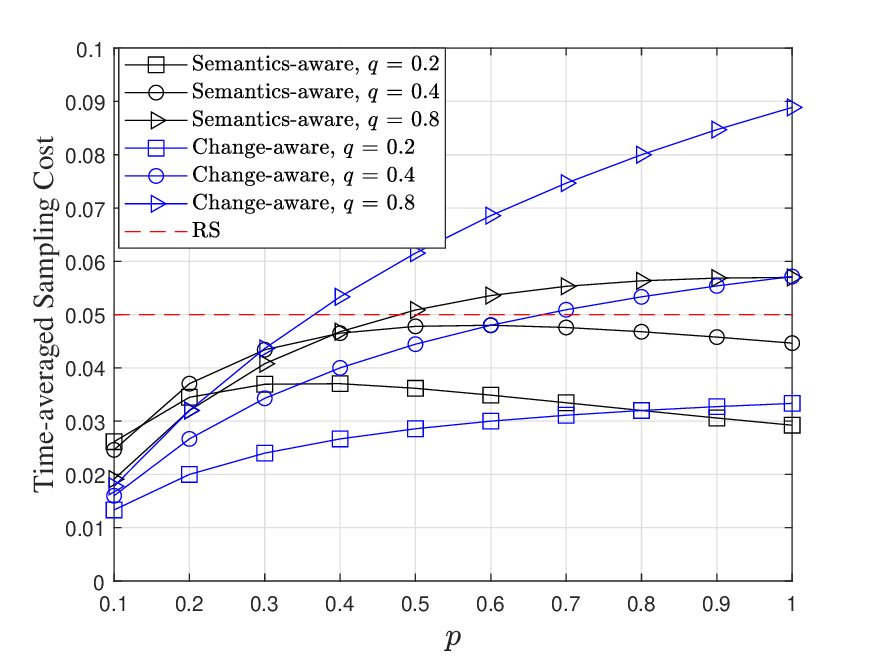}
		}
		\subfigure[$p_{{\text{s}}} = 0.7$ ]{\includegraphics[width=0.48\linewidth, clip]{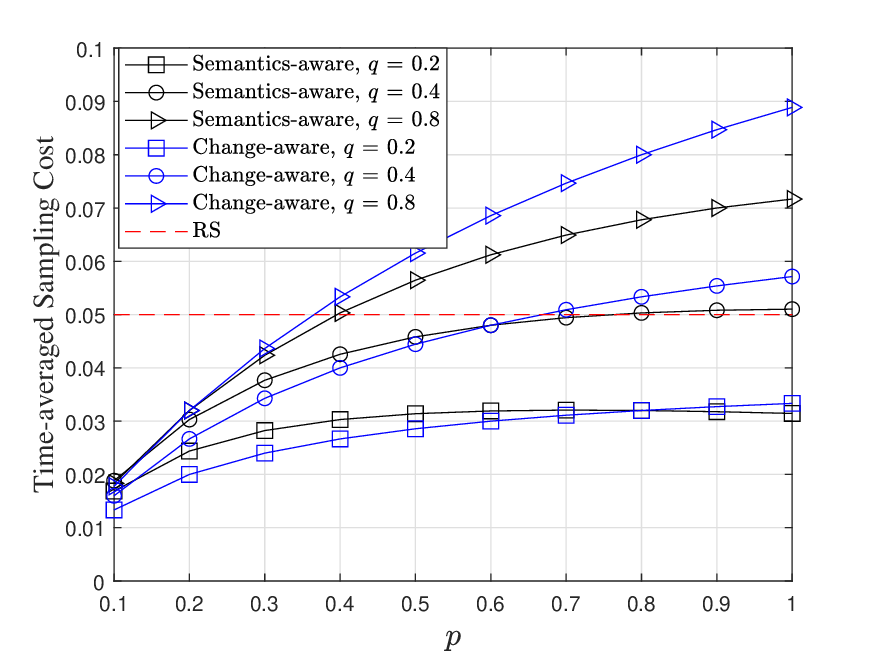}
		}
		\caption{Time-averaged sampling cost as a function of $p$ and $q$ for $\delta = 0.1$.}
		\label{Avg_SamplingCost2}
	\end{figure}
	Figs. \ref{Avg_VAoII} and \ref{Avg_SamplingCost2} illustrate the average AoIV and the time-averaged sampling cost as functions of $p$ and $q$ for $p_{\alpha^{\text{s}}}=0.5$, $\delta = 0.1$, and various $p_s$, respectively. As shown in Fig. \ref{Avg_VAoII}, the semantics-aware policy exhibits a smaller average AoIV than the other policies for slowly and rapidly changing sources. However, Fig. \ref{Avg_SamplingCost2} shows that the change-aware and semantics-aware policies generate more samples when the source changes rapidly compared to the randomized stationary policy. This implies that if there is a constraint on the sampling cost, the RSC policy outperforms the other policies for a rapidly changing source.
	
	\section{Conclusion}
	We studied a time-slotted communication system where a sampler performs sampling, and the transmitter forwards the samples over a wireless channel to a receiver that monitors the evolution of a two-state DTMC. We proposed two new metrics, namely VIA and AoIV, and we analyzed the system's average performance regarding these metrics and the AoII for the change-aware, semantics-aware, and randomized stationary policies. Furthermore, we obtained the optimal randomized stationary policy that minimizes the average VIA, subject to time-averaged sampling cost and time-averaged reconstruction error constraints. 
	
	Our results illustrated that, in terms of the average VIA, the optimal randomized stationary policy outperforms the change-aware policy for slowly and rapidly evolving sources. However, the change-aware policy exhibits better performance for moderately changing sources under certain conditions. In addition, in terms of the average AoIV, the semantics-aware policy performs the best, except under constraints on the time-averaged sampling cost and time-averaged reconstruction error are present, and when the source changes rapidly, in which cases the randomized stationary policy is superior.
	\bibliographystyle{IEEEtran}
	\bibliography{ref}
		\appendix
	\subsection{Proof of Lemma {\ref{Lemma_Pij}}}
	\label{Appendix_LemmaPij}
	\par Using the total probability theorem, we can write the transition probability given in \eqref{Pij_AoIV} as
	\begin{align}
		\label{Trans_Prob_delta_AoI}
		P_{i,j} &= \mathrm{Pr}\big[{\text{VIA}}(t+1)=j|{\text{VIA}}(t)=i\big]\notag\\ &=\mathrm{Pr}\big[{\text{VIA}}(t+1)=j|{\text{VIA}}(t)=i,X(t)=0\big]\notag\\
		&\times\mathrm{Pr}\big[X(t)=0|{\text{VIA}}(t)=i\big]\notag\\
		&+\mathrm{Pr}\big[{\text{VIA}}(t+1)=j|{\text{VIA}}(t)=i,X(t)=1\big]\notag\\
		&\times\mathrm{Pr}\big[X(t)=1|{\text{VIA}}(t)=i\big].
	\end{align}
	Now, using \eqref{Version_AoI}, one can write \eqref{Trans_Prob_delta_AoI} as
	\begin{align}
		\label{Trans_Prob_delta_AoI2}
		P_{i,j} &= \Big[\mathbbm{1}(j\!=\!0)p_{\alpha^{\text{s}}}p_{\text{s}}+\mathbbm{1}(j\!=\!i)(1-p)(1-p_{\alpha^{\text{s}}}p_{\text{s}})\notag\\
		&+\mathbbm{1}(j\!=\!i+1)p(1-p_{\alpha^{\text{s}}}p_{\text{s}})\Big]\mathrm{Pr}\big[X(t)=0|{\text{VIA}}(t)=i\big]\notag\\
		&+\Big[\mathbbm{1}(j\!=\!0)p_{\alpha^{\text{s}}}p_{\text{s}}+\mathbbm{1}(j\!=\!i)(1-q)(1-p_{\alpha^{\text{s}}}p_{\text{s}})\notag\\
		&+\mathbbm{1}(j\!=\!i+1)q(1-p_{\alpha^{\text{s}}}p_{\text{s}})\Big]\mathrm{Pr}\big[X(t)\!=\!1|{\text{VIA}}(t)\!=\!i\big],
	\end{align}
 
	where the conditional probabilities in \eqref{Trans_Prob_delta_AoI2} are given by
	\begin{align}
		\label{AoI_CondProb}
		\mathrm{Pr}\big[X(t)=0|{\text{VIA}}(t)=i\big]&=\frac{\mathrm{Pr}\big[X(t)=0,{\text{VIA}}(t)=i\big]}{\mathrm{Pr}\big[{\text{VIA}}(t)=i\big]}\notag\\&=\frac{\pi_{0,i}}{\pi_{0,i}+\pi_{1,i}},\notag\\
		\mathrm{Pr}\big[X(t)=1|{\text{VIA}}(t)=i\big]&=\frac{\mathrm{Pr}\big[X(t)=1,{\text{VIA}}(t)=i\big]}{\mathrm{Pr}\big[{\text{VIA}}(t)=i\big]}\notag\\&=\frac{\pi_{1,i}}{\pi_{0,i}+\pi_{1,i}}.
	\end{align}
	Now, using \eqref{AoI_CondProb}, we can write \eqref{Trans_Prob_delta_AoI2} as 
	
	\begin{align}
		\label{Pij_lemma_Proof}
		P_{i,j} &= \Big[\mathbbm{1}(j\!=\!0)p_{\alpha^{\text{s}}}p_{\text{s}}+\mathbbm{1}(j\!=\!i)(1-p)(1-p_{\alpha^{\text{s}}}p_{\text{s}})\notag\\
		&+\mathbbm{1}(j\!=\!i+1)p(1-p_{\alpha^{\text{s}}}p_{\text{s}})\Big]\frac{\pi_{0,i}}{\pi_{0,i}+\pi_{1,i}}\notag\\
		&+\Big[\mathbbm{1}(j\!=\!0)p_{\alpha^{\text{s}}}p_{\text{s}}+\mathbbm{1}(j\!=\!i)(1-q)(1-p_{\alpha^{\text{s}}}p_{\text{s}})\notag\\
		&+\mathbbm{1}(j\!=\!i+1)q(1-p_{\alpha^{\text{s}}}p_{\text{s}})\Big]\frac{\pi_{1,i}}{\pi_{0,i}+\pi_{1,i}}.
	\end{align}
	
	\subsection{Proof of Lemma {\ref{theorem_VAoI}}}
	\label{Appendix_LemmaPij_VAoI}
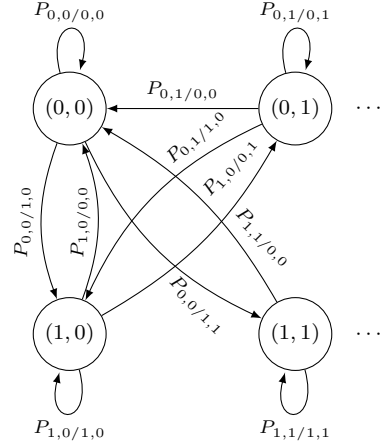
\begin{figure}[!t]
		\centering
		\begin{tikzpicture}[start chain=going left,->,>=latex,on grid,auto]
			\footnotesize
			\footnotesize
			\node[on chain]             at (3,0)(h) {$\cdots$};
			\node[state, on chain]              at (3,0) (2) {$(0,1)$};
			\node[state, on chain]             at (0,0)(1) {$(0,0)$};
			\node[state, on chain]             at (0,-3)(4) {$(1,0)$};
			\node[state, on chain]             at (3,-3)(5) {$(1,1)$};
			\node[on chain]             at (4,-3)(g) {$\cdots$};
			
			\draw[>=latex]
			(1) edge[loop above] node {\scriptsize${P_{0,0/0,0}}$}   (1)
			
			(1) edge [bend right=20, left] node [pos=0.5, left]{\rotatebox{90}{\scriptsize$P_{0,0/1,0}$}}(4)
			(1) edge[bend right=20] node [pos=0.9, left]{\rotatebox{-45}{\scriptsize$P_{0,0/1,1}$}}   (5)
			(5) edge[bend right=15] node [pos=0.1, above]{\rotatebox{-45}{\scriptsize$P_{1,1/0,0}$}}   (1)
			(2) edge[loop above] node {\scriptsize${P_{0,1/0,1}}$}   (2)
			(2) edge[above] node {\scriptsize${P_{0,1/0,0}}$}   (1)
			(2) edge[bend right=20] node [pos=0.1, left]{\rotatebox{45}{\scriptsize$P_{0,1/1,0}$}}   (4)
			(4) edge[bend right=20, right] node [pos=0.5, left]{\rotatebox{90}{\scriptsize$P_{1,0/0,0}$}}   (1)
			(4) edge[loop below] node {\scriptsize${P_{1,0/1,0}}$}   (4)
			(4) edge[bend right=15] node [pos=.95, left,yshift=-2mm]{\rotatebox{50}{\scriptsize$P_{1,0/0,1}$}}   (2)
			(5) edge[loop below] node {\scriptsize${P_{1,1/1,1}}$}   (5)
			;
		\end{tikzpicture}
		\caption{Two-dimensional DTMC describing the joint status of the original source  regarding the current state of the VIA using a two-state information source model, i.e., $\big(X(t),{\text{VIA}}(t)\big)$.}
		\label{Ver_AoI}
	\end{figure}

	To obtain $\pi_{j,i}$ we depict the two-dimensional DTMC describing the joint status of the original source  regarding the current
	state of the VIA, i.e.,  $\big(X(t), {\text{VIA}}(t)\big)$ in Fig. \ref{Ver_AoI}, where the transition probabilities $P_{i,j/m,n} = \mathrm{Pr}\big[X(t\!+\!1) \!=\! m, {\text{VIA}}(t\!+\!1)\!=\!n\big|X(t) \!=\! i,{\text{VIA}}(t)\!=\!j  \big]$, $\forall i,m\in\{0,1\}$ and $\forall j,n\in\{0,1,\cdots\}$ are given by
	
	\begin{align}
		\label{trans_prob_VAoI}
		P_{0,0/0,0}\!&=\! 1-p,\hspace{1.1cm}
		P_{0,j/0,0}\!=\! (1\!-\!p)p_{\alpha^{\text{s}}}p_{\text{s}},\notag\\
		P_{0,j/1,0} &=pp_{\alpha^{\text{s}}}p_{\text{s}},\hspace{0.6cm}
		P_{0,j/0,j+1} \!=\!0,\notag\\
		P_{0,j/1,j+1}\!&=p(1\!-\!p_{\alpha^{\text{s}}}p_{\text{s}}),\hspace{0.13cm}
		P_{0,j/0,j}=(1\!-\!p)(1\!-\!p_{\alpha^{\text{s}}}p_{\text{s}}),\notag\\
		P_{1,0/1,0} 
		&= 1-q,\hspace{1.cm}
		P_{1,j/1,0} 
		= (1-q)p_{\alpha^{\text{s}}}p_{\text{s}},\notag\\
		P_{1,j/1,j+1}&=0,\hspace{1.6cm}
		P_{1,j/0,0} =qp_{\alpha^{\text{s}}}p_{\text{s}},\notag\\
		P_{1,j/0,j+1} &=q(1-p_{\alpha^{\text{s}}}p_{\text{s}}),
		P_{1,j/1,j} =(1\!-\!q)(1\!-\!p_{\alpha^{\text{s}}}p_{\text{s}}).
	\end{align}
	Now, using Fig. \ref{Ver_AoI} and \eqref{trans_prob_VAoI}, one can derive the state stationary distribution $\pi_{j,i}$ $\forall j\in\{0,1\}$ and $i\geqslant 0$ as follows
	\begin{align}
		\!\!\pi_{0,i}\!\!&=\!\!\frac{p^kq^wp_{\alpha^{\text{s}}}p_{s}\big(1\!-\!p_{\alpha^{\text{s}}}p_{s}\big)^i}{(p\!+\!q)\big(p\!+\!(1\!-\!p)p_{\alpha^{\text{s}}}p_{s}\big)^w\big(q\!+\!(1\!-\!q)p_{\alpha^{\text{s}}}p_{s}\big)^k},
		i\!=\!0,1,\cdots.\notag\\
		\!\!\pi_{1,i}\!\!&=\!\!\frac{p^wq^kp_{\alpha^{\text{s}}}p_{s}\big(1\!-\!p_{\alpha^{\text{s}}}p_{s}\big)^i}{(p\!+\!q)\big(p\!+\!(1\!-\!p)p_{\alpha^{\text{s}}}p_{s}\big)^k\big(q\!+\!(1\!-\!q)p_{\alpha^{\text{s}}}p_{s}\big)^w}, i\!=\!0,1,\cdots.\label{RandStat_piij_VAoI}
	\end{align}
	where $k$, and $w$ are given by
	\begin{align}
		\label{kw_AoI2}
		k=
		\begin{cases}
			\frac{i}{2}, &\mod\{i,2\}=0,\\
			\frac{i+1}{2}, &\mod\{i,2\}\neq 0.
		\end{cases}\notag\\
		\hspace{0.2cm}w=
		\begin{cases}
			\frac{i+2}{2}, &\mod\{i,2\}=0,\\
			\frac{i+1}{2}, &\mod\{i,2\}\neq 0.
		\end{cases}
	\end{align}
	Similarly, for the change-aware policy $\pi_{0,i}$ and $\pi_{1,i}$ in \eqref{RandStat_piij_VAoI} can be written as
	\begin{align}
		\label{pi_CA_AoI}
		\pi_{0,i} &= \frac{qp_{\text{s}}(1-p_{\text{s}})^i}{p+q}, \hspace{0.2cm} i =0, 1, \cdots.\notag\\
		\pi_{1,i} &= \frac{pp_{\text{s}}(1-p_{\text{s}})^i}{p+q}, \hspace{0.2cm} i =0, 1, \cdots.
	\end{align}
	Using \eqref{pi_CA_AoI}, we can obtain the average VAoI, $\overline{\text{VIA}}$, for the change-aware policy as
	\begin{align}
		\label{Avg_AoIV_CA}
		\overline{\text{VIA}} = \sum_{i=0}^{\infty} i\mathrm{Pr}[{\text{VIA}}(t)=i] =  \sum_{i=0}^{\infty} i\big(\pi_{0,i}+\pi_{1,i}\big)=\frac{1-p_{\text{s}}}{p_{\text{s}}}.
	\end{align}
	\subsection{Proof of Lemma {\ref{piijk_VAoII_RS}}}
	\label{Appendix_piijk_VAoII_RS}
	\begin{figure}[!t]
		\centering
		\begin{tikzpicture}[start chain=going left,->,>=latex,on grid,auto]
			\footnotesize
			\footnotesize
			\node[state, on chain]              at (2.5,0) (2) {$(0,1,1)$};
			\node[state, on chain]             at (0,0)(1) {$(0,0,0)$};
			\node[state, on chain]             at (0,-2.5)(3) {$(1,0,1)$};
			\node[state, on chain]             at (3.5,-2.5)(4) {$(1,1,0)$};
			
			\draw[>=latex]
			(1) edge[loop above] node {\scriptsize${1-p}$}   (1)
			
			(1) edge [bend right=20, left] node [pos=0.5, left]{{\scriptsize$p(1-p_{\alpha^{\text{s}}}p_{\text{s}})$}}(3)
			(1) edge[bend right=15] node [pos=0.6, left]{\rotatebox{-45}{\scriptsize$pp_{\alpha^{\text{s}}}p_{\text{s}}$}}   (4)
			
			(2) edge[loop above] node {\scriptsize${(1-p)(1-p_{\alpha^{\text{s}}}p_{\text{s}})}$}   (2)
			(2) edge[above] node {\scriptsize${(1-p)p_{\alpha^{\text{s}}}p_{\text{s}}}$}   (1)
			(2) edge[bend right=20] node [pos=0.5, left]{{\scriptsize$p$}}   (4)
			(3) edge[bend right=20, right] node [pos=0.5, left]{{\scriptsize$q$}}   (1)
			(3) edge[loop below] node {\scriptsize${(1-q)(1-p_{\alpha^{\text{s}}}p_{\text{s}})}$}   (3)
			(3) edge[below] node {\scriptsize${(1-q)p_{\alpha^{\text{s}}}p_{\text{s}}}$}  (4)
			(4) edge[loop below] node {\scriptsize${1-q}$}   (4)
			(4) edge[bend right=20] node [pos=0.5, right]{{\scriptsize$q(1-p_{\alpha^{\text{s}}}p_{\text{s}})$}}   (2)
			(4) edge[bend right=15] node [pos=0.4, left]{\rotatebox{-45}{\scriptsize$qp_{\alpha^{\text{s}}}p_{\text{s}}$}}   (1)
			;
		\end{tikzpicture}
		\caption{Three-dimensional DTMC describing the joint status of the original and reconstructed source regarding the current state of the AoIV, i.e., $\big(X(t),\hat{X}(t),{\text{AoIV}}(t)\big)$.}
		\label{3DMC_VAoII}
	\end{figure}
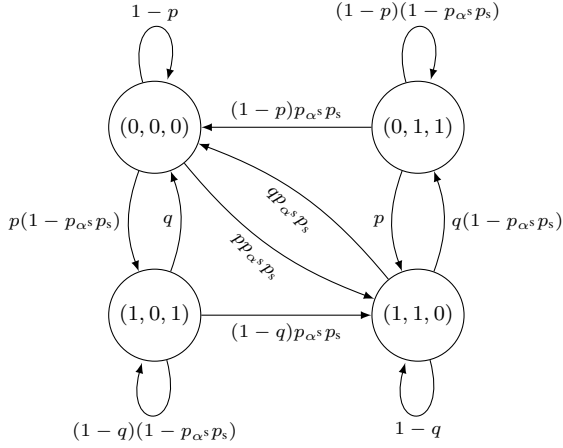
	\par To derive $\pi_{i,j,k}, \forall 
	i,j,k \in\{0,1\}$, we represent the three-dimensional DTMC describing the joint status of the original and reconstructed source regarding the current state of the AoIV, i.e., $\big(X(t),\hat{X}(t),{\text{AoIV}}(t)\big)$ as Fig. \ref{3DMC_VAoII}. Now, using Fig. \ref{3DMC_VAoII}, we can obtain the state stationary $\pi_{i,j,k}$ for the randomized stationary policy as
	
	\begin{subequations}
		\label{VAoII_pi_ijk_RS_proof}
		\begin{align}
			\pi_{0,0,0} &= \mathrm{Pr}\big[X(t)=0,\hat{X}(t)=0,{\text{AoIV}}(t)=0\big]\notag\\
			&=\frac{q\big[q+(1-q)p_{\alpha^{\text{s}}}p_{s}\big]}{(p+q)\big[p+q+(1-p-q)p_{\alpha^{\text{s}}}p_{s}\big]},\label{pi000}\\
			\pi_{0,1,1} &= \mathrm{Pr}\big[X(t)=0,\hat{X}(t)=1,{\text{AoIV}}(t)=1\big]\notag\\
			&=\frac{pq\big(1-p_{\alpha^{\text{s}}}p_{s}\big)}{(p+q)\big[p+q+(1-p-q)p_{\alpha^{\text{s}}}p_{s}\big]},\label{pi011}\\
			\pi_{1,1,0} &= \mathrm{Pr}\big[X(t)=1,\hat{X}(t)=1,{\text{AoIV}}(t)=0\big]\notag\\
			&=\frac{p\big[p+(1-p)p_{\alpha^{\text{s}}}p_{s}\big]}{(p+q)\big[p+q+(1-p-q)p_{\alpha^{\text{s}}}p_{s}\big]},\label{pi110}\\
			\pi_{1,0,1} &= \mathrm{Pr}\big[X(t)=1,\hat{X}(t)=0,{\text{AoIV}}(t)=1\big]\notag\\
			&=\frac{pq\big(1-p_{\alpha^{\text{s}}}p_{s}\big)}{(p+q)\big[p+q+(1-p-q)p_{\alpha^{\text{s}}}p_{s}\big]}\label{pi101},\\
			\pi_{0,0,1}&=\pi_{0,1,0}=\pi_{1,0,0}=\pi_{1,1,1}=0.
		\end{align}
	\end{subequations}
	Furthermore, for the change-aware policy one can write \eqref{VAoII_pi_ijk_RS_proof} as follows
	\begin{subequations}
		\label{VAoII_pi_ijk_CA_proof}
		\begin{align}
			\pi_{0,0,0} &= \frac{q}{(p+q)(2-p_{\text{s}})},
			\pi_{0,1,1} = \frac{q(1-p_{\text{s}})}{(p+q)(2-p_{\text{s}})},\label{pi011_CA}\\
			\pi_{1,1,0} &= \frac{p}{(p+q)(2-p_{\text{s}})},
			\pi_{1,0,1} = \frac{p(1-p_{\text{s}})}{(p+q)(2-p_{\text{s}})}\label{pi101_CA},\\
			\pi_{0,0,1}&=\pi_{0,1,0}=\pi_{1,0,0}=\pi_{1,1,1}=0.
		\end{align}
	\end{subequations}
	Now, using \eqref{VAoII_pi_ijk_CA_proof}, the average AoIV for the change-aware policy is given by
	\begin{align}
		\label{Avg_AoIVI_CA}
		\overline{\text{AoIV}} &= \sum_{i=1}^{\infty} i\mathrm{Pr}[{\text{AoIV}}(t)=i]=\pi_{0,1,1}+\pi_{1,0,1}=  \frac{1-p_{\text{s}}}{2-p_{\text{s}}}.
	\end{align}
	Moreover, for the semantics-aware policy, we can obtain \eqref{VAoII_pi_ijk_CA_proof} as follows
	\begin{subequations}
		\label{VAoII_pi_ijk_SA_proof}
		\begin{align}
			\pi_{0,0,0} &=
			\frac{q\big[q+(1-q)p_{s}\big]}{(p+q)\big[p+q+(1-p-q)p_{s}\big]},\label{pi000_SA}\\
			\pi_{0,1,1} &= \frac{pq\big[1-p_{s}\big]}{(p+q)\big[p+q+(1-p-q)p_{s}\big]},\label{pi011_SA}\\
			\pi_{1,1,0} 
			&=\frac{p\big[p+(1-p)p_{s}\big]}{(p+q)\big[p+q+(1-p-q)p_{s}\big]},\label{pi110_SA}\\
			\pi_{1,0,1} &= \frac{pq\big[1-p_{s}\big]}{(p+q)\big[p+q+(1-p-q)p_{s}\big]}\label{pi101_SA},\\
			\pi_{0,0,1}&=\pi_{0,1,0}=\pi_{1,0,0}=\pi_{1,1,1}=0.
		\end{align}
	\end{subequations}
	Using \eqref{VAoII_pi_ijk_SA_proof}, the average AoIV for the semantics-aware policy can be written as
	\begin{align}
		\label{Avg_AoIVI_SA}
		\overline{\text{AoIV}}&= 
		\sum_{i=1}^{\infty} i\mathrm{Pr}[{\text{AoIV}}(t)=i]=\pi_{0,1,1}+\pi_{1,0,1}\notag\\
		&=\frac{2pq\big(1-p_{s}\big)}{(p+q)\big[p+q+(1-p-\!q)p_{s}\big]}.
	\end{align}
	
	\subsection{Proof of Lemma {\ref{AoIIDistribution_RS}}}
	\label{Appendix_AoIIDistribution_RS}
	\par
\begin{figure}[!t]
		\centering
		\scriptsize
		\begin{tikzpicture}[start chain=going left,->,>=latex,node distance=2.5cm]
			\node[state]    (A)                     {$(0,0)$};
			\node[state]    (B)[above right of=A]   {$(0,1)$};
			\node[state]    (C)[below right of=A]   {$(1,0)$};
			\node[state]    (D)[below right of=B]   {$(1,1)$};
			\path
			(A) edge[loop left]     node{$1-p$}  (A)
			edge[bend left=15,above]  node{$pp_{\alpha^{\text{s}}}p_{\text{s}}$}  (D)
			edge[bend right=15,left]    node{$p(1-p_{\alpha^{\text{s}}}p_{\text{s}})$}  (C)
			(B) edge[loop above]  node{$(1-p)(1-p_{\alpha^{\text{s}}}p_{\text{s}})$}     (B)
			edge[left=15] node{$(1-p)p_{\alpha^{\text{s}}}p_{\text{s}}$}    (A)
			edge[bend right=15,left] node{$p$}    (D)
			(C) edge[loop below]  node{$(1-q)(1-p_{\alpha^{\text{s}}}p_{\text{s}})$}     (C)
			edge[bend right=15,right] node{$q$}    (A)
			edge[ left=15,right] node{$(1-q)p_{\alpha^{\text{s}}}p_{\text{s}}$}    (D)
			(D) edge[loop right]    node{$1-q$}     (D)
			edge[bend right=15,right]  node{$q(1-p_{\alpha^{\text{s}}}p_{\text{s}})$}  (B)
			edge[bend left=15,above]     node{$qp_{\alpha^{\text{s}}}p_{\text{s}}$}         (A);
		\end{tikzpicture}
		\vspace*{1ex}
		\caption{Two-dimensional DTMC describing the joint status of the original source regarding the current state at the reconstructed source using a two-state information source model.}
		\label{2DimMarkovChain}
	\end{figure}
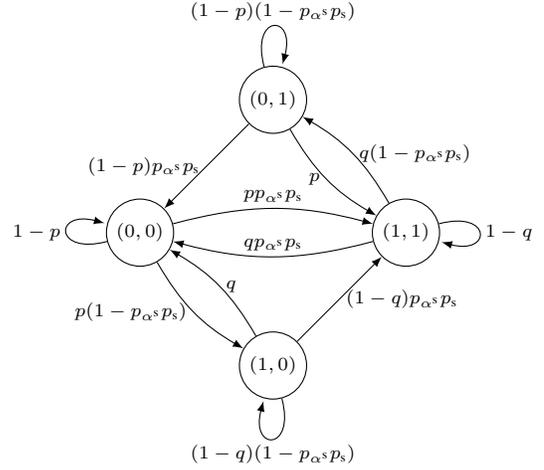

 At time slot $t$, $\text{AoII}(t)$ denotes the synced state, therefore $\mathrm{Pr}\big[{\text{AoII}}(t)=0\big]$ can be written as
	\begin{align}
		\label{AoII_Sync}
		\mathrm{Pr}\big[{\text{AoII}}(t)=0\big] &= \mathrm{Pr}\big[X(t)=0, \hat{X}(t)=0\big]\notag\\
		&+\mathrm{Pr}\big[X(t)=1, \hat{X}(t)=1\big] = \pi^{\text{A}}_{0,0}+\pi^{\text{A}}_{1,1}.
	\end{align}
	Furthermore,  $\text{AoII}(t)=i\geqslant 1$ indicates that the system was in a synced state at time slot $t-i$, and it has been in an erroneous state from time slots $t-i+1$ to $t$. Therefore, we can calculate $\mathrm{Pr}\big[\text{AoII}(t)=i\big]$ as follows
	\begin{align}
		\label{AoII_NoSync}
		&\mathrm{Pr}\big[{\text{AoII}}(t)\!=\!i\big]\notag\\
		&= \mathrm{Pr}\big[X(t)\!\neq\! \hat{X}(t), X(t\!-\!1)\!\neq\! \hat{X}(t\!-\!1), \cdots\notag\\
		&, X(t\!-\!i\!+\!1)\neq \hat{X}(t\!-\!i\!+\!1),X(t\!-\!i)= \hat{X}(t\!-\!i)\big]\notag\\
		&=\mathrm{Pr}\big[X(t)\!\neq\! \hat{X}(t), X(t\!-\!1)\!\neq\! \hat{X}(t\!-\!1), \cdots\notag\\
		&, X(t\!-\!i\!+\!1)\neq \hat{X}(t\!-\!i\!+\!1)\big|X(t\!-\!i)= 0,\hat{X}(t\!-\!i)=0\big]\notag\\
		&\times\mathrm{Pr}\big[X(t\!-\!i)= 0,\hat{X}(t\!-\!i)=0\big]+\mathrm{Pr}\big[X(t)\!\neq\! \hat{X}(t),\cdots\notag\\
		&, X(t\!-\!i\!+\!1)\neq \hat{X}(t\!-\!i\!+\!1)\big|X(t\!-\!i)= 1,\hat{X}(t\!-\!i)=1\big]\notag\\
		&\times\mathrm{Pr}\big[X(t\!-\!i)= 1,\hat{X}(t\!-\!i)=1\big]\notag\\
		&=p(1-q)^{i-1}(1\!-\!p_{\alpha^{\text{s}}}p_{\text{s}})^{i}\pi^{\text{A}}_{0,0}\!+\!q(1-p)^{i-1}(1\!-\!p_{\alpha^{\text{s}}}p_{\text{s}})^{i}\pi^{\text{A}}_{1,1},
	\end{align}
	where $\pi^{\text{A}}_{i,j}, \forall i,j\in\{0,1\}$ are the probabilities derived from the stationary distribution of the two-dimensional DTMC describing the joint status of the originl source regarding the current state at the reconstructed source, i.e., $\big(X(t), \hat{X}(t)\big)$. Now, using the two-dimensional DTMC depicted in Fig. \ref{2DimMarkovChain}, one can obtain $\pi^{\text{A}}_{i,j}$ as follows
	\begin{align}
		\label{pij_2D_RS}
		\pi^{\text{A}}_{0,0} &=\frac{q\big[q+(1-q)p_{\alpha^{\text{s}}}p_{\text{s}}\big]}{(p+q)\big[p(1-p_{\alpha^{\text{s}}}p_{\text{s}})\!+\!q\!+\!(1-q)p_{\alpha^{\text{s}}}p_{\text{s}}\big]},\notag\\\hspace{0.1cm}
		\pi^{\text{A}}_{0,1} &=\frac{pq(1-p_{\alpha^{\text{s}}}p_{\text{s}})}{(p+q)\big[p(1-p_{\alpha^{\text{s}}}p_{\text{s}})\!+\!q\!+\!(1-q)p_{\alpha^{\text{s}}}p_{\text{s}}\big]},\notag\\
		\pi^{\text{A}}_{1,0} &=\frac{pq(1-p_{\alpha^{\text{s}}}p_{\text{s}})}{(p+q)\big[p(1-p_{\alpha^{\text{s}}}p_{\text{s}})\!+\!q\!+\!(1-q)p_{\alpha^{\text{s}}}p_{\text{s}}\big]},\hspace{0.1cm}\notag\\
		\pi^{\text{A}}_{1,1} &=\frac{p\big[p+(1-p)p_{\alpha^{\text{s}}}p_{\text{s}}\big]}{(p+q)\big[p(1-p_{\alpha^{\text{s}}}p_{\text{s}})\!+\!q\!+\!(1-q)p_{\alpha^{\text{s}}}p_{\text{s}}\big]}.
	\end{align}
	Now, using \eqref{pij_2D_RS}, $\mathrm{Pr}\big[\text{AoII}(t)=i\big]$ given in \eqref{AoII_Sync} and \eqref{AoII_NoSync} can be written as
	\begin{align}
		\label{PrAoII_RS}
		&\mathrm{Pr}\big[\text{AoII}(t)=i\big] \notag\\
		&= 
		\begin{cases}
			\frac{p^2+q^2+(p+q-p^2-q^2)p_{\alpha^{\text{s}}}p_{\text{s}}}{(p+q)\big[p+q+(1-p-q)p_{\alpha^{\text{s}}}p_{\text{s}}\big]},\hspace{0.2cm} &i=0,\\
			\frac{pq(1-p_{\alpha^{\text{s}}}p_{\text{s}})^{i}\big[(1-q)^{i-1}\Phi(q)+(1-p)^{i-1}\Phi(p)\big]}{(p+q)\big[p+q+(1-p-q)p_{\alpha^{\text{s}}}p_{\text{s}}\big]},\hspace{0.2cm} &i\geqslant 1.
		\end{cases}
	\end{align}
	where $\Phi(x)=x+(1-x)p_{\alpha^{\text{s}}}p_{\text{s}}$.
	Similarly, for the change-aware policy, $\mathrm{Pr}\big[\text{AoII}(t)=i\big]$ in \eqref{PrAoII_RS} is given by
	\begin{align}
		\label{PrAoII_CA}
		\!\!\!\mathrm{Pr}\big[\text{AoII}(t)=i\big] \!=\! 
		\begin{cases}
			\frac{1}{2-p_{\text{s}}},\hspace{0.2cm} &i=0,\\
			\frac{pq(1-p_{\text{s}})\big[(1-p)(1-q)\big]^{i-1}}{(p+q)(2-p_{\text{s}})},\hspace{0.2cm} &i\geqslant 1.
		\end{cases}
	\end{align}
	Now, using \eqref{PrAoII_CA}, the average AoII for the change-aware policy is obtained as
	\begin{align}
		\label{Avg_AoII_CA}
		\overline{\text{AoII}} &=\sum_{i=1}^{\infty} i\mathrm{Pr}[\text{AoII}(t)\!=\!i]=\frac{(p^2+q^2)(1-p_{\text{s}})}{pq(p+q)(2-p_{\text{s}})}.
	\end{align}
	Furthermore, for the semantics-aware policy, we can obtain $\mathrm{Pr}\big[\text{AoII}(t)=i\big]$ as follows
	\begin{align}
		\label{PrAoII_SA}
		&\mathrm{Pr}\big[\text{AoII}(t)=i\big] \notag\\
		&= 
		\begin{cases}
			\frac{p^2+q^2+(p+q-p^2-q^2)p_{\text{s}}}{(p+q)\big[p+q+(1-p-q)p_{\text{s}}\big]},\hspace{0.2cm} &i=0,\\
			\frac{pq(1-p_{\text{s}})^{i}\big[(1-q)^{i-1}\Psi(q)+(1-p)^{i-1}\Psi(p)\big]}{(p+q)\big[p+q+(1-p-q)p_{\text{s}}\big]},\hspace{0.2cm} &i\geqslant 1.
		\end{cases}
	\end{align}
	where $\Phi(x)=x+(1-x)p_{\alpha^{\text{s}}}p_{\text{s}}$.
	Now, using \eqref{PrAoII_SA}, one can calculate the average AoII for the semantics-aware policy as
	\begin{align}
		\label{Avg_AoII_SA}
		\overline{\text{AoII}} &=\sum_{i=1}^{\infty} i\mathrm{Pr}[\text{AoII}(t)\!=\!i]\notag\\ &=\frac{pq(1-p_{s})\big[p+q+(2-p-q)p_{s}\big]}{(p+q)\Psi(p)\Psi(q)\big[p+q+(1-p-q)p_{s}\big]}.
	\end{align}
	where $\Psi(x)=x+(1-x)p_{s}$.
\end{document}